\documentclass[footinbib,aip,prl,reprint]{revtex4-2}

\usepackage{graphicx}
\usepackage{amsmath}
\usepackage{mathtools}
\usepackage{amssymb}
\usepackage{xcolor}

\usepackage{graphicx}
\usepackage{bm}

\begin{document}


\title{Nearly-Resonant Crystalline-Phononic Coupling in Quantum Spin Liquid Candidate CsYbSe$_2$}

\author{Yun-Yi Pai}
\email{yunyip@ornl.gov}
\address{Materials Science and Technology Division, Oak Ridge National Laboratory, Oak Ridge, TN 37831}
\address{Quantum Science Center, Oak Ridge, Tennessee 37831}

\author{Claire E. Marvinney}
\address{Materials Science and Technology Division, Oak Ridge National Laboratory, Oak Ridge, TN 37831}
\address{Quantum Science Center, Oak Ridge, Tennessee 37831}

\author{Liangbo Liang}
\address{Center for Nanophase Materials Sciences, Oak Ridge National Laboratory, Oak Ridge, TN 37831}

\author{Jie Xing}
\address{Materials Science and Technology Division, Oak Ridge National Laboratory, Oak Ridge, TN 37831}

\author{Allen Scheie}
\address{Neutron Scattering Division, Oak Ridge National Laboratory, Oak Ridge, TN 37831}

\author{Alexander A. Puretzky}
\address{Center for Nanophase Materials Sciences, Oak Ridge National Laboratory, Oak Ridge, TN 37831}

\author{G\'{a}bor B. Hal\'{a}sz}
\address{Materials Science and Technology Division, Oak Ridge National Laboratory, Oak Ridge, TN 37831}
\address{Quantum Science Center, Oak Ridge, Tennessee 37831}

\author{Xun Li}
\address{Materials Science and Technology Division, Oak Ridge National Laboratory, Oak Ridge, TN 37831}

\author{Rinkle Juneja}
\address{Materials Science and Technology Division, Oak Ridge National Laboratory, Oak Ridge, TN 37831}

\author{Athena S. Sefat}
\address{Materials Science and Technology Division, Oak Ridge National Laboratory, Oak Ridge, TN 37831}

\author{David Parker}
\address{Materials Science and Technology Division, Oak Ridge National Laboratory, Oak Ridge, TN 37831}

\author{Lucas Lindsay}
\address{Materials Science and Technology Division, Oak Ridge National Laboratory, Oak Ridge, TN 37831}

\author{Benjamin J. Lawrie}
\email{lawriebj@ornl.gov; This manuscript has been authored by UT-Battelle, LLC, under contract DE-AC05-00OR22725 with the US Department of Energy (DOE). The US government retains and the publisher, by accepting the article for publication, acknowledges that the US government retains a nonexclusive, paid-up, irrevocable, worldwide license to publish or reproduce the published form of this manuscript, or allow others to do so, for US government purposes. DOE will provide public access to these results of federally sponsored research in accordance with the DOE Public Access Plan (http://energy.gov/downloads/doe-public-access-plan). }
\address{Materials Science and Technology Division, Oak Ridge National Laboratory, Oak Ridge, TN 37831}
\address{Quantum Science Center, Oak Ridge, Tennessee 37831}

\date{\today}

\begin{abstract}
CsYbSe$_2$, a recently identified quantum spin liquid (QSL) candidate, exhibits strong crystal electric field (CEF) excitations. Here, we identify phonon and CEF modes with Raman spectroscopy and observe strong CEF-phonon mixing resulting in a vibronic bound state. Complex, mesoscale interplay between phonon modes and CEF modes is observed in real space, and an unexpected nearly resonant condition is satisfied, yielding up to fifth-order combination modes, with a total of 17 modes identified in the spectra. This study paves the way to coherent control of possible QSL ground states with optically accessible CEF-phonon manifolds and mesoscale engineering of CEF-phonon interactions.  
\end{abstract}


\maketitle

\section{Introduction}

Quantum spin liquids (QSLs) have been a topic of intense research interest in condensed matter physics in recent decades \cite{ANDERSON1973153,Knolle2019,Savary2016}. The still elusive possibility of control over distributed many-body entanglement offers a key path toward fault-tolerant registers for quantum information processing. Quantum spin liquid candidates like $\alpha$-RuCl$_3$ \cite{Kasahara2018,Sandilands2015,WangYiping2020,Wulferding2020,Banerjee2016NM,Banerjee1055,Baek2017,Jansa2018,pai2021magnetostriction}, $\kappa$-(BEDT-TTF)$_2$Cu$_2$(CN)$_3$ \cite{doi:10.1126/science.abc6363}, and YbMgGaO$_4$ \cite{RN3956YbMgGaO4} typically exhibit geometric frustration of some sort, but there have not yet been conclusive observations of QSLs, nor are there well accepted measurement protocols for QSL order parameters.

Yb delafossites\cite{Schmidt_2021} like NaYbS$_2$ \cite{NaYbS2_2018,NaYbS2_2019}, NaYbSe$_2$ \cite{NaYbSe2PRB_CEF,NaYbSe2_2,NaYbSe2_PRX}, NaYbO$_2$ \cite{NaYbO2_1,NaYbO2_2,NaYbO2_3}, CsYbSe$_2$\cite{Xing_Jie_PRB2019Rapid,pocs2021systematic,xie2021fieldinduced,RN3955}, and KYbSe$_2$ \cite{xing2021synthesis,scheie2021witnessing} - with effective spin $S_{\text{eff}} = 1/2$ from Yb$^{3+}$ ions and antiferromagnetic coupling decorating 2D triangular lattices - are popular candidates for the realization of quantum spin liquids. Easy-plane magnetic anisotropy results in a plateau at one-third of the saturation magnetization due to strong spin fluctuations\cite{NaYbSe2_2, Xing_Jie_PRB2019Rapid, NaYbO2_1, NaYbO2_3}. There have been no reports of long-range magnetic ordering at temperatures as low as $T = $ 0.4 K as a result of the geometric frustration in these materials. This family is thought to be less defect prone than QSL candidate YbMgGaO$_4$\cite{a_large_family,RN3955,xie2021fieldinduced,Xing_Jie_PRB2019Rapid}. 

Crystal electric field (CEF) modes emerge as a result of the lifting of the orbital degeneracy of transition metal ions in certain environments\cite{HansBethe, VanVleck}. CEFs can therefore be used as probes of internal fields. Lower energy CEFs are particularly important as they are part of the ground state wavefunction that defines properties such as magnetism, conductivity and superconductivity\cite{Schaack_Book_2000}. 
For instance, the CEF splitting and spin-orbit coupling of Yb$^{3+}$ 4f$^{14}$ orbitals causes the intrinsic magnetic anisotropy in Yb delafossites. CEF modes are also responsible for a Schottky anomaly that has been observed in some members of this family. The rare earth Yb$^{3+}$ doublets are time-reversal degenerate \cite{kramers1930theorie}. The 4f electrons - well shielded compared to 3d electrons - exhibit smaller CEF splitting \cite{Lanthanide_CEF_intro}. CEF excitations can also interact with other excitations in the system\cite{Sethi_PRL_2019_omegas,k6695,adroja2012vibron}. While CEFs have been extensively studied since the 60s, understanding of their interactions with other excitations beyond pristine single ion environments remains rather limited. CEF coupling to electron-hole pairs has been proposed as the mechanism that mediates superconductivity in UPd$_2$Al$_3$\cite{UPD_SC1, UPD_SC2, UPD_SC3, UPD_SC4} and PrOs$_4$Sb$_{12}$\cite{PrOs4Sb12_SC}. It was reported for Pr(Os$_{1-x}$Ru$_{x}$)$_{4}$Sb$_{12}$ that, when CEF and phonon modes cross each other, the superconducting transition temperature has a minimum\cite{mini_Tc_when_CEF_phonon_cross}.

Near-resonant interactions between CEF and phonon modes can give rise to strong CEF-phonon coupling, which has been observed in CeAuAl$_3$ \cite{k6695, CeAuAl}, Ce$_2$O$_3$\cite{Sethi_PRL_2019_omegas}, CeAl$_2$ \cite{JTCeAl2, RamanCeAl2, Thalmeier_1982}, PrNi$_5$ \cite{CEFPhononPrNi5} NdBa$_2$Cu$_3$O$_{7-\delta}$\cite{CEF_highTC}, and Ho$_2$Ti$_2$O$_7$\cite{Gaudet_CEF_ph_Ho2Ti2O7}, among other materials. Conduction electrons may further complicate the picture and alter the relative intensity of hybrid modes\cite{CEFEatenConfigCrossOver}. Understanding the complex manifold spanned by CEFs and phonons, their interactions, and their associated optical selection rules may lead to emergent functionalities that take advantage of spin-lattice coupling. Also, this understanding may be useful for controlling and reading out possible QSL ground states for topological quantum information processing. 

Inelastic scattering methods like neutron and Raman scattering are workhorse tools for probing fundamental excitations such as phonons, CEF modes \cite{Klein1983}, magnons, and possible Majorana states \cite{Wulferding2020, HighfieldSahasrabudhe}. Though Raman spectroscopy typically probes excitations near the $\Gamma$ point in the Brillouin zone, it is capable of probing higher energy modes that are challenging to access with neutron scattering. Further, Raman microscopy can be used to probe sub-micron-scale spatial variation to reveal interactions between excitations. Here, we employ polarization-, spatially-, and temperature-resolved Raman spectroscopy to probe CEF-phonon interactions in CsYbSe$_2$. We identify all the primary phonons, and CEFs. We report combination modes, a vibronic bound state, and mode repulsion between CEF and phonon modes. Our work extends explorations of CEF-phonon coupling into the emerging class of candidate QSLs.

\begin{figure}
\centering
    \includegraphics[width=1\columnwidth]{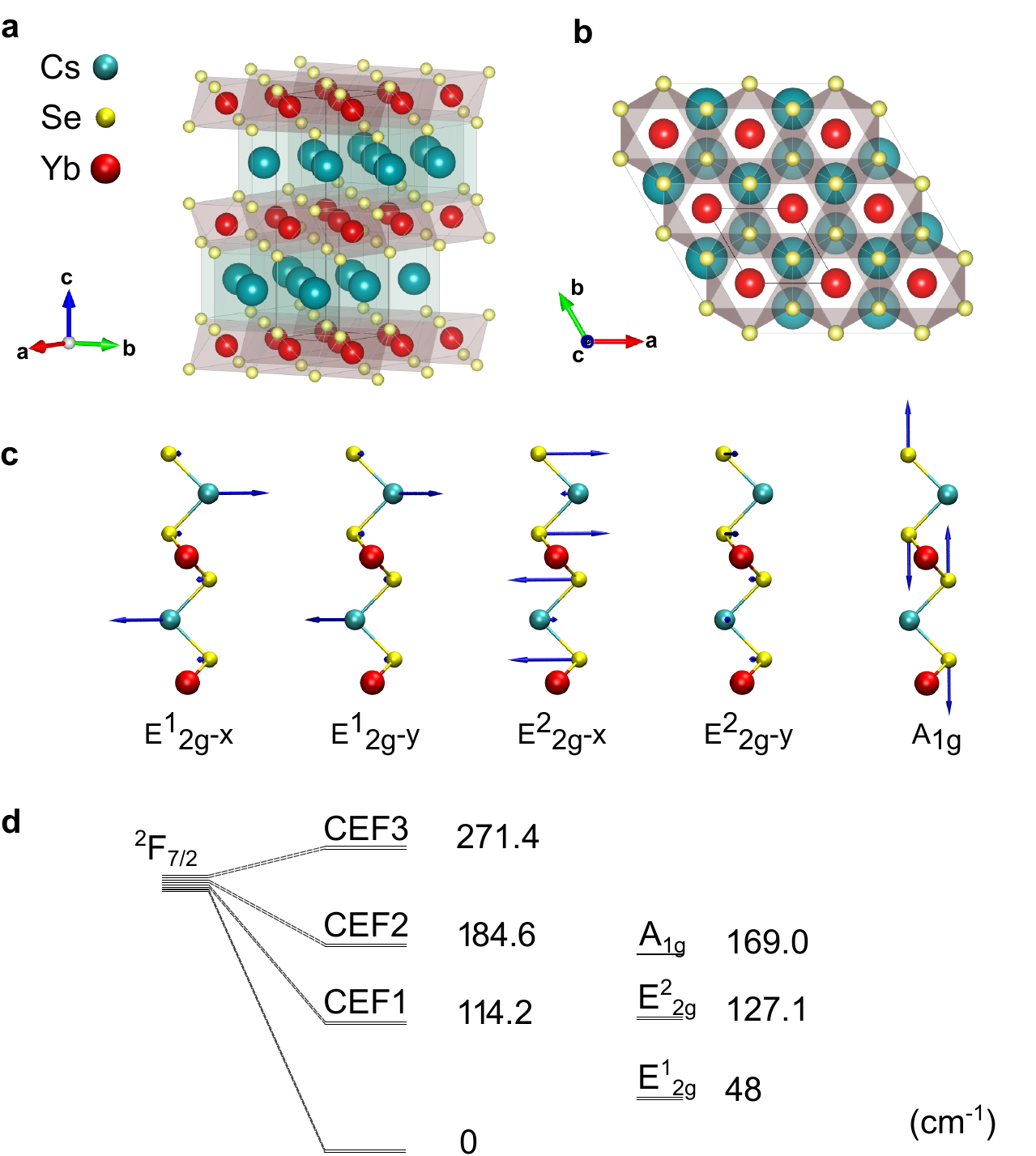}
    \caption{(a) Side view and (b) top view of the crystal structure of CsYbSe$_2$. (c) Vibration patterns of Raman active phonon modes in CsYbSe$_2$. (d) Primary CEF excitations and phonon excitations. } 
    \label{fig:crystal}
\end{figure}

\section{Results and Discussion}

\subsection{Theoretical analysis of phonon and CEF modes}
Figures \ref{fig:crystal} (a) and (b) show the crystal structure of CsYbSe$_2$. According to group theory and our first-principles phonon calculations, CsYbSe$_2$ belongs to the space group $P$6$_3$/$mmc$ (No. 194) with the point group D$_{\text{6h}}$. The bulk unit cell has a hexagonal lattice with 8 atoms in total, giving rise to 24 normal phonon modes at the $\Gamma$ point: $\Gamma$(D$_{\text{6h}}$) = A$_{\text{1g}}$ + 3A$_{\text{2u}}$ + 2B$_{\text{1u}}$ + 2B$_{\text{2g}}$ + E$_{\text{1g}}$ + 2E$_{\text{2g}}$ + 3E$_{\text{1u}}$ + 2E$_{\text{2u}}$, where Raman active phonon modes correspond to non-degenerate A$_{\text{1g}}$ symmetry modes, doubly degenerate E$_{\text{2g}}$ symmetry modes, and doubly degenerate E$_{\text{1g}}$ symmetry modes. The intensity of a Raman mode is $I \propto | \mathbf{e}_s \cdot \widetilde{R} \cdot \mathbf{e}_i^T|^2$, where $\widetilde{R}$ is the Raman tensor of a phonon mode, and $\mathbf{e}_s$ and $\mathbf{e}_i$ are the electric polarization vectors of the scattered and incident light, respectively\cite{MoS2_WSe2_Shear_breathing, Liangbo_Fe3GeTe2}. In the back-scattering geometry with linear laser polarization, $\mathbf{e}_s$ and $\mathbf{e}_i$ are in plane. Based on the Raman tensors shown in the Supplemental Information, it is clear that only A$_{\text{1g}}$ and E$_{\text{2g}}$ phonon modes can appear in the Raman spectra, whereas E$_{\text{1g}}$ modes cannot be observed in the our experimental configuration. This is common in hexagonal layered materials\cite{Liangbo_Fe3GeTe2}. Our DFT calculated frequencies for the two E$_{\text{2g}}$ modes (E$_{\text{2g}}^{\text{1}}$ and E$_{\text{2g}}^{\text{2}}$) are 37 cm$^{-1}$ and 97 cm$^{-1}$, and A$_{\text{1g}}$ at 122 cm$^{-1}$.  These results are consistent with our experimental Raman observations that will be discussed below.

\subsection{Temperature dependence}

\begin{figure}
\centering
    \includegraphics[width=1\columnwidth]{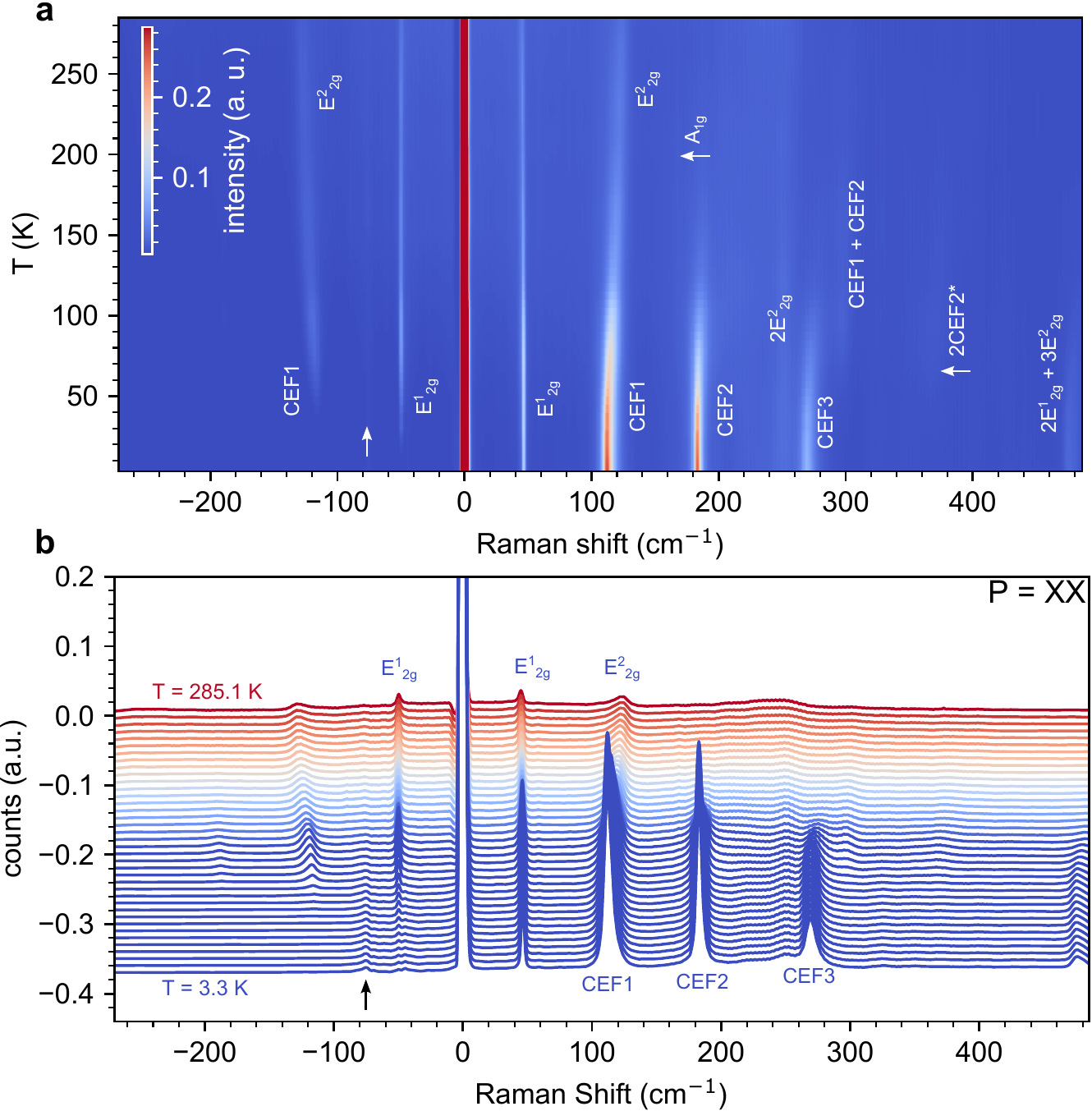}
    \caption{(a) Temperature dependent Raman spectra from $T = 3.3$ K to $T = 273$ K. The legends correspond to the assignment of the peaks. (b) The line traces of the temperature dependence data from (a). The white arrow in (a) and the black arrow in (b) indicate an artifact from volume Bragg gratings that leaked through the spectrometer.}
    \label{fig:temp}
\end{figure}

Temperature dependent Raman spectra acquired in an XX polarization configuration ($\textbf{e}_s$ = (1, 0, 0), $\textbf{e}_i$ = (1, 0, 0)) from $T = 3.3$ K to $T = 285$ K (1 K step from 3.3 to 30 K, 5K step from 40 K to 285 K) are shown in Figure \ref{fig:temp} (a) and (b). The white arrow in (a) and the black arrow in (b) highlight an artifact from the volume Bragg gratings that is unrelated to CsYbSe$_2$. 

A total of 10 Raman modes on the Stokes side are resolved. We assign the three modes at 114 cm$^{-1}$, 183 cm$^{-1}$, and 269 cm$^{-1}$ to CEF1, CEF2, and CEF3 respectively. Since the CEF excitations in CsYbSe$_2$ are expected to be dominated by Yb$^{3+}$ 4f$^{14}$ orbitals with the electronic ground state $J = 7/2$ manifold that is split into 4 doubly degenerate Kramer pairs (Figure \ref{fig:crystal} (d)), these three modes and the ground state comprise the expected 4 pairs. The next excited state manifold $J = 5/2$ is expected to be about 10,000 cm$^{-1}$ higher\cite{Koningstein_1967}. Additonally, CEFs typically exhibit strongly temperature-dependent intensity and linewidth \cite{Schaack_Book_2000}. CEF1-CEF3 soften in energy at low temperatures and becomes significantly more prominent. In particular, CEF1 is more than 15.3 times stronger in intensity at $T = 3.3$ K than at $T = 285$ K. The energy levels for the three CEF1-CEF3 modes are consistent with previous spectroscopies of CEF modes in NaYbSe$_2$ \cite{NaYbSe2PRB_CEF} and KYbSe$_2$\cite{scheie2021witnessing}. The CEF1 and CEF2 modes are also consistent with a recent report on CEF levels probed by resonant torsion magnetometry and low field susceptibility measurements that are particularly sensitive to CEF1 and the details of the CEF Hamiltonian \cite{pocs2021systematic}.

Though it is common to fit experimentally observed CEF energy levels to obtain crystal field parameters, three clearly visible CEF peaks provide too little information to constrain a fit to the six nonzero crystal field parameters for the three-fold Yb$^{3+}$ symmetry in CsYbSe$_2$. Nevertheless, to get an approximate understanding of the CEF ground state, we modeled the CEF Hamiltonian using a point charge model calculated using PyCrystalField software\cite{PyCrytalField} and the Se environment of CsYbSe$_2$. To match energy scales, we fitted the Se effective charges to the three measured low-temperature CEF levels. This yielded an effective Se charge of -1.54e for CsYbSe$_2$. The observed and calculated CEF level energies are in Table~\ref{table}. The ground states of these models, assuming a quantization axis in the c direction, are 

\begin{equation}
    |\psi_\pm \rangle  = \pm0.968|\pm7/2\rangle   +  0.218|\pm1/2\rangle \pm 0.128|\pm5/2\rangle
\end{equation}

This point charge fit is a crude approximation, and shows an easy-axis ground state. However, the coefficients for the ground state eigenkets can easily be adjusted to give an easy-plane ground state.  Previous experience with these materials \cite{pocs2021systematic, NaYbSe2PRB_CEF} shows that their ground states tend to be more isotropic or easy-planar, suggesting that the above point-charge fit is not very accurate.

\begin{figure*}[hbt!]
\centering
    \includegraphics[width=2\columnwidth]{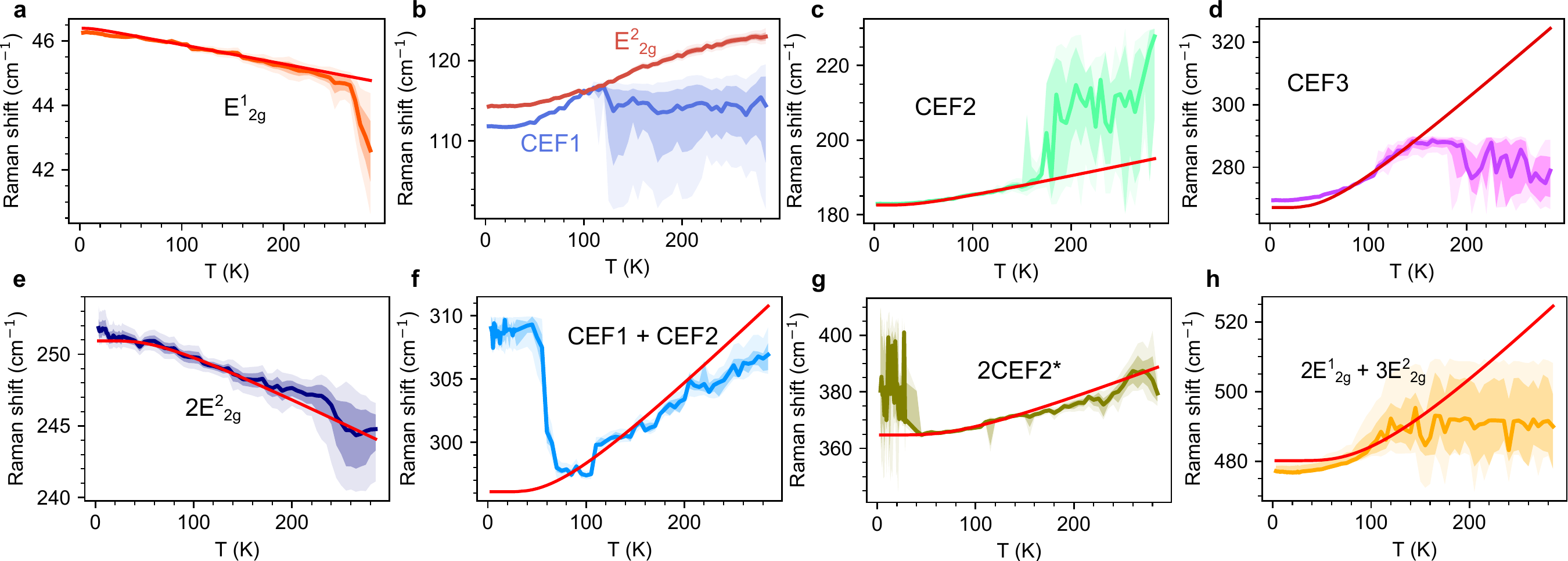}
    \caption{Peak positions as a function of temperature for the most prominent 9 peaks (a) E$^1_{\text{2g}}$, (b) CEF1 and E$^2_{\text{2g}}$, (c) CEF2, (d) CEF3, (e) 2E$^2_{\text{2g}}$, (f) CEF1 + CEF2, (g) 368.0 cm$^{-1}$ and (h) 2E$^1_{\text{2g}}$ + 3E$^2_{\text{2g}}$. The peak positions are extracted from Bayesian inference on the data in Figure \ref{fig:temp}. The shaded intervals are the 68\% HDI and 95\% HDI.}
    \label{fig:fit}
\end{figure*}

Based on our DFT phonon calculations discussed above, we assign the E$^1_{\text{2g}}$ mode to the 46 cm$^{-1}$ peak. Further, we assign the peak at 120 cm$^{-1}$ at $T = 293$ K to E$^2_{\text{2g}}$ (see Supplemental Information for linecuts at $T = 293$ K). At first glance, it may seem that this high-temperature mode transitions continuously to the peak at 114 cm$^{-1}$ at $T = 3.3$ K, which we assigned to CEF1 because of its strong softening and intensity increase at low temperatures. However, more careful analysis of the Raman spectra (discussed in greater detail below) shows that, at $T = 293$ K, CEF1 is a weak mode nearly resonant with E$^2_{\text{2g}}$. As the temperature decreases, the amplitude of  E$^2_{\text{{2g}}}$ drops compared to CEF1. The A$_{\text{1g}}$ mode is assigned to a small peak at 169.0 cm$^{-1}$, since this peak shows up in the XX polarization configuration and disappears in the XY polarization configuration, a signature of A$_{\text{1g}}$ symmetry according to the group theory analysis discussed below. While this mode is resolved at $T = 293$ K, due to its low intensity and its proximity to CEF2, we were not able to resolve it at low temperatures. We assign the mode 255.9 cm$^{-1}$ to 2E$^2_{\text{2g}}$, and 300 cm$^{-1}$ to CEF1 + CEF2. The 368.0 cm$^{-1}$ and 479.1 cm$^{-1}$ peaks could be attributed to several higher-order modes including 368.0 cm$^{-1}$:  2CEF2, CEF1 + 2E$^2_{\text{2g}}$, CEF3 + 2E$^1_{\text{2g}}$ and 479.1 cm$^{-1}$: 2E$^1_{\text{2g}}$  + 3E$^2_{\text{2g}}$, CEF1 + 2CEF2. We observed and identified 8 more modes in higher energy Raman spectra (Supplemental Information) that we will revisit in Section \ref{resonantsection}. 

To further verify the mode assignment and symmetry, we performed polarization-resolved Raman spectroscopy measurements in both parallel (XX) and cross (XY) polarization configurations. According to group theory analysis (more details in Supplemental Information), the polarization profile of an E$_{\text{2g}}$ Raman mode is circular under any linear polarization configuration. For an A$_{\text{1g}}$ Raman mode, the polarization profile is circular in the XX configuration while its intensity is zero in the XY configuration ($\textbf{e}_s$ = (1, 0, 0), $\textbf{e}_i$ = (0, 1, 0)). Similar polarization responses of Raman modes have been previously reported for NaYbSe$_2$\cite{NaYbSe2PRB_CEF}. Under the XX configuration, at room temperature,  E$^1_{\text{2g}}$, E$^2_{\text{2g}}$, and A$_{\text{1g}}$ all have 0-fold symmetry (i.e., circular polarization profiles), and at low temperatures, the E$^1_{\text{2g}}$, CEF1, CEF2 and CEF3 modes likewise exhibit 0-fold symmetry (Supplemental Information), consistent with the group theory analysis for these modes. All of the modes above except A$_{\text{1g}}$ exhibit nearly the same intensity across the XX and XY configurations while A$_{\text{1g}}$ is suppressed in the XY configuration, in agreement with the analysis based on their Raman tensors (Supplemental Information). We note that the polarization profiles of CEF Raman modes are very similar to the polarization profiles of E$_{\text{2g}}$ phonon modes, suggesting that CEF modes share similar forms of Raman tensors to E$_{\text{2g}}$ phonon modes. 

To track the temperature evolution of the observed CEF and phonon modes, we employed Bayesian inference techniques with a Hamiltonian Monte Carlo technique \texttt{PyMC3}\cite{pyMC}, a No U-Turns (NUTS) sampler, and 4 chains with 3,000 samples for spectral modeling. Figure \ref{fig:fit} illustrates the inferred peak positions of the most prominent peaks: E$^1_{\text{2g}}$, E$^2_{\text{2g}}$, CEF1, CEF2, CEF3, 2E$^2_{\text{2g}}$, CEF1 + CEF2, 368.0 cm$^{-1}$ and 479.1 cm$^{-1}$. The traces are the median values from the Bayesian inference. The two shaded bands are 68 \% (darker) and 95 \% (lighter) highest density intervals (HDIs). The red trace is a fit of the resulting trace assuming a phonon anharmonicity relationship. The E$^1_{\text{2g}}$ mode harden as temperature decreases. The modes that have CEF character all soften. The CEF modes are more prominent for $T < 120 - 140$ K and far weaker at higher temperatures. This can be seen in the large uncertainty of the Bayesian inference result for these CEF modes at higher temperatures. The mode CEF1 + CEF2 at 300 cm$^{-1}$ (Figure \ref{fig:fit} (d)) disappears at $T < 80$ K, suggesting that its initial state is a CEF excited state. Figure \ref{fig:fit} (b) shows the interacting E$^2_{\text{2g}}$ and CEF1 modes. CEF1 has small amplitude at a lower energy than E$^2_{\text{2g}}$ at higher temperatures and become 15.3 times stronger at $T = 3.3$ K while E$^2_{\text{2g}}$ becomes small relative to CEF1.

\begin{figure*}[t]
\centering
    \includegraphics[width=2\columnwidth]{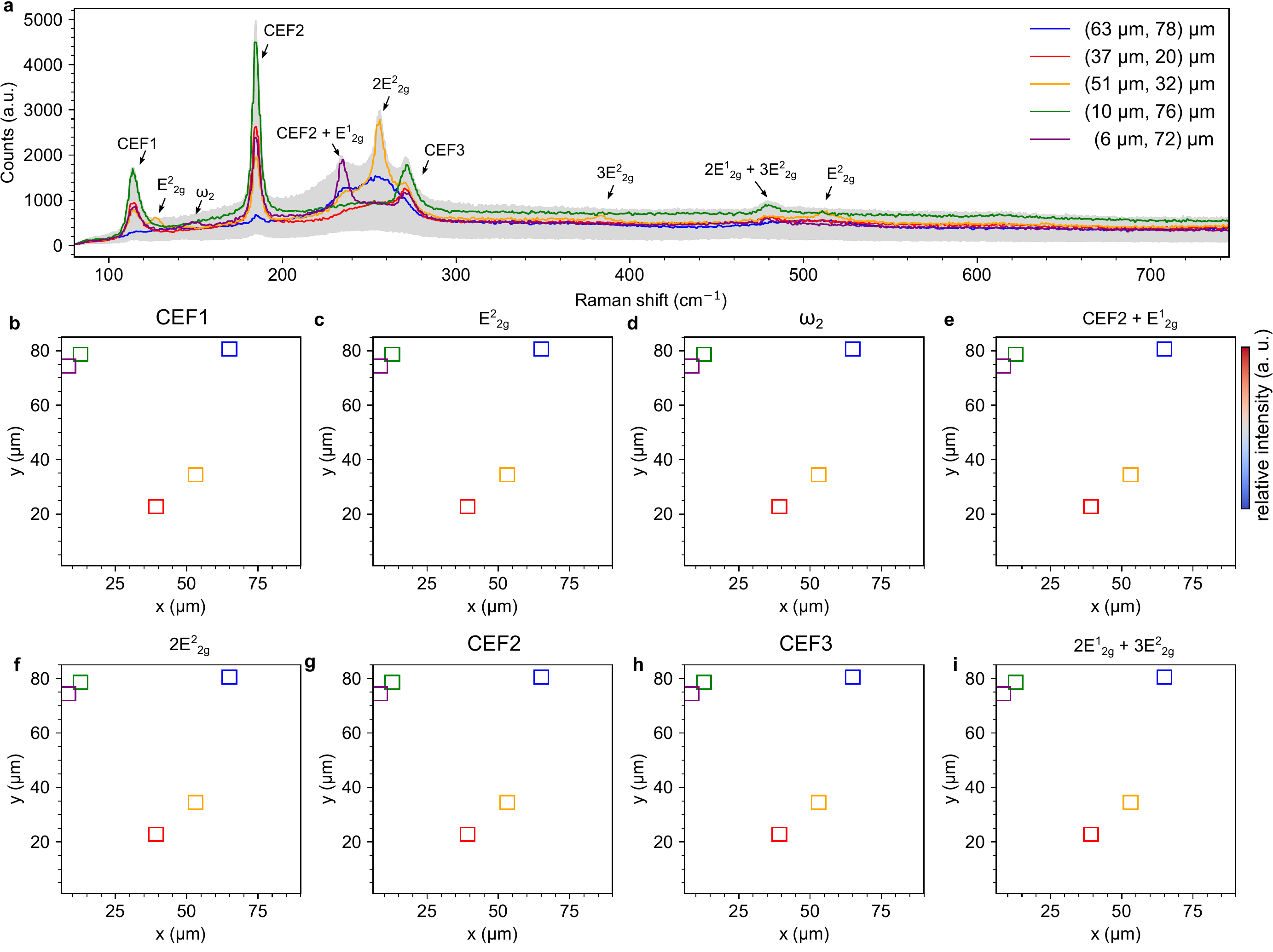}
    \caption{Hyperspectral Raman map of a 96.1 $\mu$m by 88.4 $\mu$m area of the crystal shows subtle mesoscale spatial dependence of the CEF modes and phonon modes. The data was taken with no polarization control. (a) Representative spectra at 5 positions on the sample. (b)-(i) integrated relative intensity of the prominent peaks as a function of position. (b) CEF1, (c) E$^2_{\text{2g}}$, (d) $\omega_2$, (e) CEF2 + E$^2_{\text{2g}}$, (f) 2E$^2_{\text{2g}}$, (g) CEF2, (h) CEF3, (i) 2E$^1_{\text{2g}}$ + 3E$^2_{\text{2g}}$. }
    \label{fig:spatial}
\end{figure*}

\subsection{Real space mode repulsion observed with hyperspectral Raman}
Mesoscale interplay between CEF and phonon modes is also observed in real-space hyperspectral Raman microscopy of the same flake at $T = 3.3 $ K, as shown in Figure \ref{fig:spatial}. The Raman map consists of 45 $\times$ 45 spectra across a 96 $\times$ 88 $\mu$m area. Selected spectra with distinct representative features are shown in Figure \ref{fig:spatial} (a). 

The spectrum at $(x, y) =(10 \;\mu\text{m}, 76 \;\mu\text{m})$ has prominent CEF1, CEF2, and CEF3 modes, consistent with the majority of the 2,025 sampled pixels. While the average of all the spectra (illustrated by the shaded background) still exhibits most of the spectral content that is present at each pixel, material heterogeneity induces substantial linewidth broadening in the shaded spectrum. For example, at $(x, y) =(6 \;\mu\text{m}, 72\;\mu\text{m})$, the intensity of the CEF1, CEF2 and CEF3 modes are only about 50 \% of their peak intensity, and smaller peaks at  127.1 cm$^{-1}$ and 145.1 cm$^{-1}$ are observed. Additionally, the peaks at 233.9 cm$^{-1}$ and 255.9 cm$^{-1}$ are more prominent. There are also smaller peaks at 383.7 cm$^{-1}$ and 510.7 cm$^{-1}$. We assign 127.1 cm$^{-1}$ to E$^2_{\text{2g}}$, which is only a small baseline next to CEF1 at most pixels and 145.1 cm$^{-1}$ to a CEF-phonon coupling mode $\omega_2$. We further assign 255.9 cm$^{-1}$, 383.7 cm$^{-1}$ and 510.7 cm$^{-1}$ to 2E$^2_{\text{2g}}$, 3E$^2_{\text{2g}}$ and 4E$^2_{\text{2g}}$, respectively. 

To carefully track the spatial dependence of each mode we utilized baseline removal with asymmetric least squares (ALS) and non-negative matrix factorization (NMF). Below we discuss the ALS results. NMF results are described in the Supplementary Information. 

Figures \ref{fig:spatial} (b) - (i) illustrate the Raman intensity maps (red: high; blue: low) for modes (b) CEF1, (c) E$^2_{\text{2g}}$, (d) $\omega_2$, (e) CEF2 + E$^1_{\text{2g}}$ (f) 2E$^2_{\text{2g}}$, (g) CEF2, (h) CEF3 and (i) 2E$^1_{\text{2g}}$ + 3E$^2_{\text{2g}}$. The selected pixels from panel (a) are marked. Spatial anticorrelations between the CEF modes and E$^2_{\text{2g}}$ phonons are observed: where CEFs are strong, the E$^2_{\text{2g}}$ and  2E$^2_{\text{2g}}$ modes are weak. While there have been many spectroscopic reports of CEF-phonon coupling in other materials, interplay of CEF excitations and phonons in real space has not been previously reported. One possible origin of the inhomogeneity may come from Yb atoms occupying Cs sites (as reported for NaYbSe$_2$ \cite{NaYbSe2_PRX}). A rigorous description of the origin of mesoscale phonon-CEF interactions is beyond the scope of this paper, but the ability to control phonon-CEF interactions, through defect engineering, for example, could enable substantial advances in the development of quantum technologies based on Yb delafossite QSLs.

\subsection{CEF-phonon bound state and combination modes}\label{resonantsection}
\textbf{Bound state}. When CEF and phonon modes are nearly resonant, vibronic bound states can form. This form of \textit{magnetoelastic} coupling has been detected by Raman spectroscopy since the 80s in CeAl$_2$\cite{Thalmeier_1982}. There, closely spaced CEF and phonon modes form two new modes with energies described by the Thalmeier-Fulde description \cite{Thalmeier_1982}:
\begin{equation} 
\omega_{\text{1,2}} = \frac{\omega_{\text{CEF}} + \omega_{\text{ph}} }{2} \pm \sqrt{ (\frac{\omega_{\text{CEF}} - \omega_{\text{ph}} }{2})^2 + V^2   }
\end{equation}
where $\omega_{\text{CEF}}$ and $\omega_{\text{ph}}$ are the energies of the closely spaced CEF and phonon modes, respectively, and $V$ is the effective coupling strength. With the 145.1 cm$^{-1}$ peak assigned to the $\omega_2$ mode of CEF1 and E$^2_{\text{2g}}$, we obtain an effective coupling strength of 23.6 cm$^{-1}$ at $T = 3.3$ K, which is smaller than Ce$_2$O$_3$\cite{Sethi_PRL_2019_omegas} and larger than CeAl$_2$ \cite{Thalmeier_1982}. This model suggests that a conjugate bound state $\omega_1$ exists at 96.2 cm$^{-1}$, but $\omega_1$ was not observed, potentially due to the 90 cm$^{-1}$ cutoff of the longpass filter that was used for the hyperspectral Raman measurements. It is worth pointing out that $\omega_2$ is observed only in a small subset of the sampled real space positions, such as $(x, y) =(6 \;\mu\text{m}, 72\;\mu\text{m})$, the purple trace in Figure \ref{fig:spatial} (a) and purple square in the rest of the subplots of Figure \ref{fig:spatial}. The eigenvibration of the  E$^2_{\text{2g}}$ mode is described by the out-of-phase vibration of the two Se atoms next to the Yb$^{3+}$ ion, which is responsible for the CEF modes. One might ask how the CEF-phonon coupling of these two modes alters the ground state wavefunction and how the lower energy excitations accommodate this change. While the nascent literature argues that CsYbSe$_2$ is less defect prone than other QSL candidates \cite{RN3955,xie2021fieldinduced,Xing_Jie_PRB2019Rapid}, defects are still responsible for many of the observed properties. For example, the sister material NaYbSe$_2$ was reported to have 4.8\% of Na sites occupied by Yb ions\cite{NaYbSe2_PRX}. A similar effect is likely present in CsYbSe$_2$ as Na and Cs both have the same valence as Yb. However, the presence of Yb atoms at Cs sites yielding a stronger phonon response and weaker CEF is counterinuitive as Yb is the cause of the CEF, so one would expect the CEF to be stronger at the defect sites. Furthermore, the E$_{\text{2g}}$ modes are mainly due to the vibrations of Cs atoms. Heavier Yb (173.04 u) atoms at Cs (132.91 u) sites should yield a weaker Raman response. 

\textbf{Combination modes}. Together, a total of 17 modes are observed in Figures \ref{fig:temp}, \ref{fig:spatial} and S1. Their energies and assignments are summarized in Table \ref{table}. Up to fifth order combination modes such as CEF2 + E$^1_{\text{2g}}$ + 3E$^2_{\text{2g}}$ and 2CEF2 + CEF3, 5E$^2_{\text{2g}}$ are identified. Phonon dressed electronic excitations are extremely common in photolumninescence\cite{polaronic_CrI3} and ARPES\cite{polaron_vdW}, and combination modes and overtones are also quite common in resonant Raman spectroscopy, including CEF modes coupled to LO phonons\cite{CEF_multiLO} and CEF subtraction modes \cite{CEF_subtraction}. However, overtones and combination modes are far less common in non-resonant Raman spectroscopy. Resonant absorption with 532 nm (2.33 eV) excitation is not expected in this system, as there is no known absorption line for CsYbSe$_2$ near 2.33 eV. 

\begin{table}
\begin{tabular}{ c c  } 
 wavenumber (cm$^{-1})$  & assignment   \\ 
 \hline
 48.0 & E$^1_{\text{2g}}$   \\ 
 \hline
96.2 &$\omega_1$ from CEF1 and E$^2_{\text{2g}}$ (V = 23.5839)  \\ 
 \hline
114.2 &CEF1  \\ 
 \hline
127.1 &E$^2_{\text{2g}}$   \\ 
 \hline
145.1 &$\omega_2$ from CEF1 and E$^2_{\text{2g}}$ (V = 23.5839)    \\ 
 \hline
169.0 &A$_{1g} (293\; \text{K}) $   \\ 
 \hline
184.6 &CEF2   \\ 
 \hline
 233.9 &CEF2 +  E$^1_{\text{2g}}$   \\ 
 \hline
255.9 &2E$^2_{\text{2g}}$   \\ 
 \hline
271.4 &CEF3   \\ 
 \hline
300.0 &CEF1 + CEF2    \\ 
 \hline
368.0 &2CEF2, CEF1 + 2E$^2_{\text{2g}}$, CEF3 + 2E$^1_{\text{2g}}$\\ 
 \hline
383.7 & 3E$^2_{\text{2g}}$, CEF1 + CEF3   \\ 
 \hline
479.1 & 2E$^1_{\text{2g}}$  + 3E$^2_{\text{2g}}$, CEF1 + 2CEF2  \\ 
 \hline
510.7 & 4E$^2_{\text{2g}}$   \\ 
 \hline
615.4 & CEF2 + E$^1_{\text{2g}}$ + 3E$^2_{\text{2g}}$, 3CEF1 + CEF3 \\ 
 \hline
639.6 &2CEF2 + CEF3, 5E$^2_{\text{2g}}$    \\ 
\end{tabular}
\caption{\label{tab:table-name}Summary of assignments of the observed modes}\label{table}
\end{table}

\section{Conclusion}
We identified all the primary CEF and phonon modes for QSL candidate CsYbSe$_2$ and verified their symmetries. Interestingly, with sub-micron spot size, we observed mesoscale spatial mode repulsion between CEF and E$^2_{\text{2g}}$ (and 2E$^2_{\text{2g}}$) phonon modes. Furthermore, we identified a CEF-phonon bound state $\omega_2$ from CEF1 and E$^2_{\text{2g}}$ and extracted an effective coupling strength V = 23.58 cm$^{-1}$. These results for the magnetoelastic coupling in CsYbSe$_2$ can be used to estimate the coupling between phonons and potential spinons which may enable the confirmation of the underlying QSL ground state\cite{Phonon_Kitaev, Phonon_Kitaev2}. Understanding the mechanism behind the mesoscale CEF-phonon coupling with knowledge of the complex CEF and phonon manifolds may provide a pathway to optically addressable mesoscale quantum spin devices that take advantage of the QSL ground state.

\section{Methods}

\subsection{Sample Details and Experimental Setup}
High quality single crystal CsYbSe$_2$ (Figure \ref{fig:crystal}) was grown using a previously described flux method \cite{RN3955}. Polarization-resolved Raman spectra from $T = 3.3$ K - 300 K were taken in a Montana Instruments closed-cycle cryostat with out-of-plane excitation and a back scattering geometry (beam path $\|$ \textbf{c}). The spectra were measured with a Princeton Instruments Isoplane SCT-320 spectrograph with a Pixis 400BR Excelon camera and a 2400 line/mm grating. A 532.03 nm continuous wave laser excitation and a set of 3 Optigrate volume Bragg gratings were used to access low energy Stokes and anti-Stokes Raman modes. Achromatic half-wave plates were mounted on piezoelectric rotators for polarization control. The power at the sample was about 1.5 mW and the typical acquisition time was about 30 sec per spectrum. The hyperspectral Raman microscopy was performed with an AttoCube 3-axis positioner and Semrock filters in lieu of the Bragg gratings (for improved collection efficiency).  

\subsection{Calculation Details}
To obtain frequencies and vibration patterns of Raman-active phonon modes in CsYbSe$_2$, first-principles plane-wave density functional theory (DFT) calculations were performed using VASP software with projector augmented wave (PAW) pseudopotentials for electron-ion interactions\cite{KRESSE_1996_15} and the Perdew-Burke-Ernzerhof (PBE) functional for exchange-correlation interactions\cite{PBE}. The DFT+U method was used to consider the localized f electrons of Yb atoms, with the effective U parameter chosen as 6.0 eV\cite{Dudarev_1998}. Other U values including 4.0, 5.0, and 7.0 eV were tested as well, which yielded the same phonon frequencies. Both atomic coordinates and lattice constants were optimized until the residual forces were below 0.001 eV/\AA, with a cutoff energy of 400 eV and a $\Gamma$-centered k-point sampling of 18$\times$18$\times$5. The total energy was converged to 10$^{-8}$ eV. Based on the optimized unit cell, we then performed phonon calculations using a finite-difference scheme implemented in Phonopy\cite{phonopy}. The same cutoff energy and convergence criteria of energy were used. Hellmann-Feynman forces in the 3$\times$3$\times$1 supercell
with a $\Gamma$-centered k-point sampling of 6$\times$6$\times$5 were computed by VASP for both positive and negative atomic displacements ($\delta$ = 0.03 \AA) and then used in Phonopy to construct the dynamical matrix. The diagonalization of the dynamical matrix provides phonon frequencies and eigenvectors (calculation results in Supplemental Information). 

\acknowledgments
This research was sponsored by the U. S. Department of Energy, Office of Science, Basic Energy Sciences, Materials Sciences and Engineering Division. The first-principles phonon calculations and Raman microscopy were performed at the Center for Nanophase Materials Sciences, which is a U.S. Department of Energy Office of Science User Facility. L.L. acknowledges computational resources of the Compute and Data Environment for Science (CADES) at the Oak Ridge National Laboratory, which is supported by the Office of Science of the U.S. Department of Energy under Contract No. DE-AC05-00OR22725. Postdoctoral research support was provided by the Intelligence Community Postdoctoral Research Fellowship Program at the Oak Ridge National Laboratory, administered by Oak Ridge Institute for Science and Education through an interagency agreement between the U.S. Department of Energy and the Office of the Director of National Intelligence.

\section{Author contributions}
All authors discussed the results thoroughly. Y.-Y. P., C. E. M., and B. J. L. performed most of the measurements. L. L. performed Raman tensor analysis. J. X. grew the samples. A. S. performed crystal field analysis. Y.-Y. P, L. L. did majority of the data analysis with inputs from A. A. P.. X. L., R. J., L. L. performed DFT phonon calculations. A. S. S., D. P., L. L. and B. J. L initiated and oversaw the project. Y.-Y. P., L.L., B. J. L wrote most of the manuscript with contributions from all authors.

\bibliography{references}

\end{document}


\maketitle

\section{Temperature Dependence for Higher Energy Band}
\begin{figure}[htbp]
\centering
\includegraphics[width=1.0\linewidth]{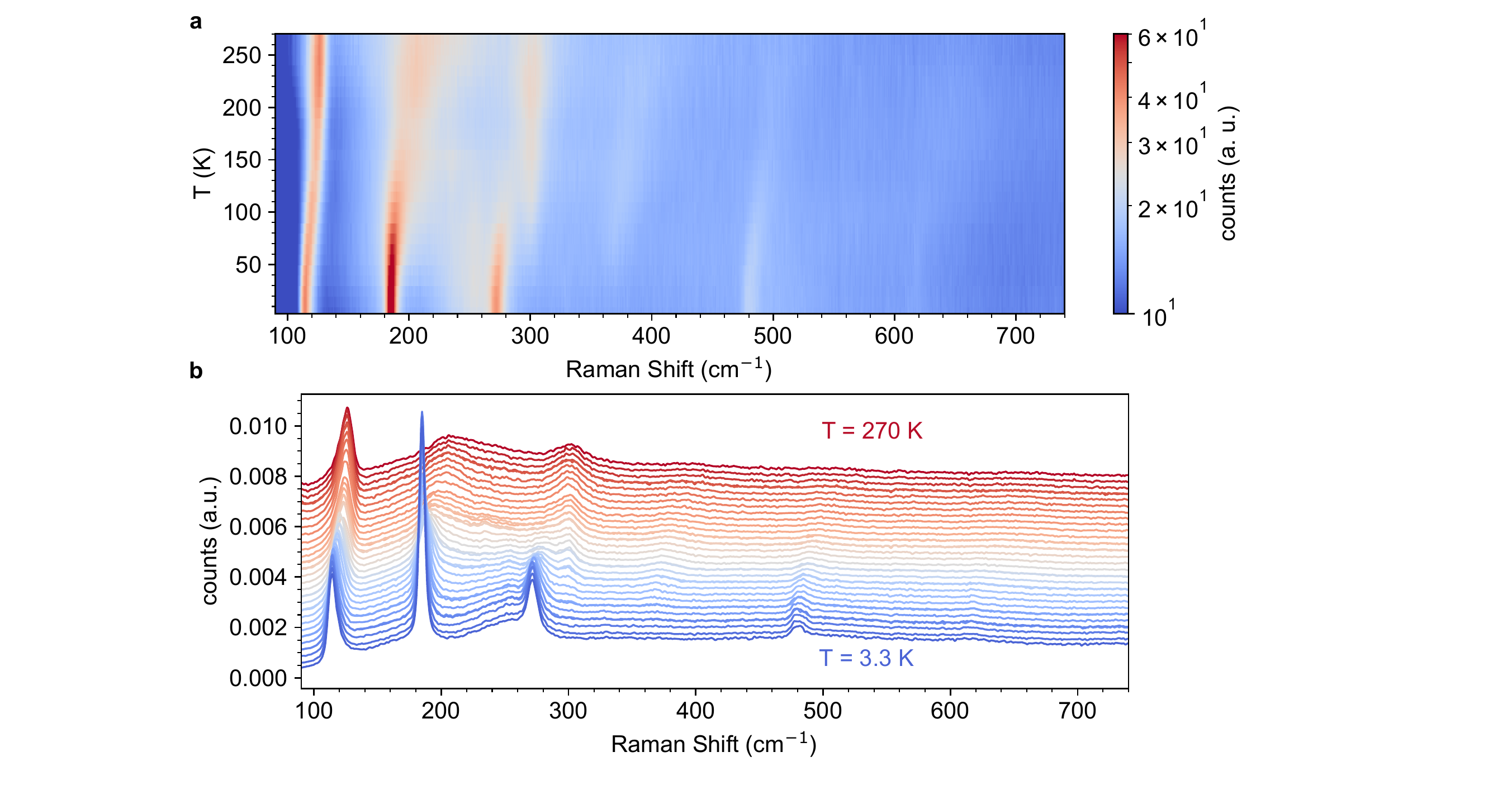}
\caption{Raman spectra in a higher energy band as a function of temperature from $T =$ 3.3 K to $T =$ 270 K.}
\label{fig:highenergy}
\end{figure}

Figure \ref{fig:highenergy} shows temperature-dependent unpolarized Raman spectra taken in a higher energy band from $T =$ 3.3 K to $T =$ 270 K. The spectra were taken with Semrock dichroic and longpass filters instead of a set of volume Bragg gratings.

\newpage

\section{Raman group theory analysis for CsYbSe$_2$}
CsYbSe$_2$ belongs to the space group $P$6$_3$/$mmc$ (No. 194) with the point group D$_{\text{6h}}$, and its Raman active phonon modes consist of non-degenerate A$_{\text{1g}}$ symmetry modes, doubly degenerate E$_{\text{2g}}$ symmetry modes (E$_{\text{2g}}-x$ and E$_{\text{2g}}-y$), and doubly degenerate E$_{\text{1g}}$ symmetry modes (E$_{\text{1g}}-x$ and E$_{\text{1g}}-y$). Their Raman tensors $\widetilde{R}$ assume the following forms:

\begin{equation*} \widetilde{R} (\text{A}_{\text{1g}})  = 
\begin{pmatrix}
a & \cdot & \cdot\\
\cdot & a & \cdot \\
\cdot & \cdot & b \\
\end{pmatrix}
\end{equation*}

\begin{equation*} 
\widetilde{R} (\text{E}_{\text{2g}}-x)  = 
\begin{pmatrix}
c     & \cdot & \cdot\\
\cdot & -c    & \cdot \\
\cdot & \cdot & \cdot \\
\end{pmatrix}; 
\widetilde{R} (\text{E}_{\text{2g}}-y)  = 
\begin{pmatrix}
\cdot & c     & \cdot\\
c     & \cdot & \cdot \\
\cdot & \cdot & \cdot \\
\end{pmatrix}
\end{equation*}

\begin{equation} 
\widetilde{R} (\text{E}_{\text{1g}}-x)  = 
\begin{pmatrix}
\cdot & \cdot & \cdot\\
\cdot & \cdot & d     \\
\cdot & d     & \cdot \\
\end{pmatrix}; 
\widetilde{R} (\text{E}_{\text{1g}}-y)  = 
\begin{pmatrix}
\cdot & \cdot & d    \\
\cdot & \cdot & \cdot \\
d     & \cdot & \cdot \\
\end{pmatrix}
\label{eq:Raman_tensor}
\end{equation}

{\setlength{\parindent}{0cm}In the experimental back-scattering laser geometry (light $Z$ in and $Z$ out) with linear polarization, the electric polarization vectors of the scattered and incident light $\bold{e}_s$ and $\bold{e}_i$ are in-plane (i.e., the $X$-$Y$ plane), and they are given by $\bold{e}_s$ = (cos$\gamma$, sin$\gamma$, 0) and $\bold{e}_i$ = (cos$\theta$,  sin$\theta$, 0). With Raman intensity $I \propto | \bold{e}_s \cdot \widetilde{R} \cdot \bold{e}_i^T|^2$ , we have: }
\begin{equation}
 I \propto \left|\begin{pmatrix} cos\gamma, & sin\gamma, & 0\end{pmatrix}\cdot \widetilde{R} \cdot \begin{pmatrix} cos\theta\\ sin\theta\\ 0 \end{pmatrix}\right|^2
 \label{eq:int_general}
\end{equation}

{\setlength{\parindent}{0cm} It is obvious that E$_{\text{1g}}$ phonon modes have zero Raman intensity in the back-scattering geometry, and thus cannot be observed experimentally. For A$_{\text{1g}}$ and E$_{\text{2g}}$ modes, by substituting the Raman tensors $\widetilde{R}$ from Eq. \ref{eq:Raman_tensor} into Eq. \ref{eq:int_general}, we can obtain}

\begin{align}
 I(\text{A}_{\text{1g}}) &\propto \left|\begin{pmatrix}  cos\gamma, & sin\gamma, & 0\end{pmatrix} \cdot \begin{pmatrix}
a & \cdot & \cdot\\
\cdot & a & \cdot \\
\cdot & \cdot & b \\
\end{pmatrix} \cdot \begin{pmatrix} cos\theta\\ sin\theta\\ 0 \end{pmatrix}\right|^2   \nonumber \\
           &\propto \left|a cos\gamma cos \theta + a sin\gamma sin \theta \right|^2  \nonumber \\
           &\propto |a|^2 cos^2(\gamma - \theta)  \label{eq:final_a1g}
\end{align}
\begin{align}
 I(\text{E}_{\text{2g}}-x) &\propto \left|\begin{pmatrix}  cos\gamma, &  sin\gamma, & 0\end{pmatrix} \cdot \begin{pmatrix}
c     & \cdot & \cdot\\
\cdot & -c    & \cdot \\
\cdot & \cdot & \cdot \\
\end{pmatrix} \cdot \begin{pmatrix} cos\theta\\ sin\theta\\ 0 \end{pmatrix}\right|^2 \nonumber \\
 &\propto \left|\begin{pmatrix}  c cos\gamma, &  -c sin\gamma, & 0\end{pmatrix} \cdot \begin{pmatrix} cos\theta\\ sin\theta\\ 0 \end{pmatrix}\right|^2  \nonumber \\
 &\propto \left|c cos\gamma cos \theta -c sin\gamma sin \theta \right|^2  \nonumber \\
 &\propto |c|^2 cos^2(\gamma + \theta)  
\end{align}
\begin{align}
 I(\text{E}_{\text{2g}}-y) &\propto \left|\begin{pmatrix}  cos\gamma, &  sin\gamma, & 0\end{pmatrix} \cdot \begin{pmatrix}
\cdot    & c & \cdot\\
c & \cdot   & \cdot \\
\cdot & \cdot & \cdot \\
\end{pmatrix} \cdot \begin{pmatrix} cos\theta\\ sin\theta\\ 0 \end{pmatrix}\right|^2 \nonumber \\
 &\propto \left|\begin{pmatrix}  c sin\gamma, &  c cos\gamma, & 0\end{pmatrix} \cdot  \begin{pmatrix} cos\theta\\ sin\theta\\ 0 \end{pmatrix}\right|^2  \nonumber \\
 &\propto \left|c sin\gamma cos \theta + c cos\gamma sin \theta \right|^2  \nonumber \\
 &\propto |c|^2 sin^2(\gamma + \theta)
\end{align}
Consequently, the Raman intensity of a doubly degenerate E$_{\text{2g}}$ mode is 
\begin{equation}
 I(\text{E}_{\text{2g}})  = I(\text{E}_{\text{2g}}-x) + I(\text{E}_{\text{2g}}-y) = |c|^2 cos^2(\gamma + \theta) + |c|^2 sin^2(\gamma + \theta) = |c|^2
\label{eq:final_E_2g}
\end{equation}
Eq. \ref{eq:final_E_2g} indicates that the polarization profile of an E$_{\text{2g}}$  phonon mode is a circle under any linear polarization configuration. For an A$_{\text{1g}}$  phonon mode, under the experimental parallel polarization configuration (i.e., XX, $\gamma = \theta $), its intensity I(A$_{\text{1g}}$) $\propto$  $|a|^2$ according to Eq. \ref{eq:final_a1g} and hence the polarization profile is also a circle; however, under the experimental cross polarization configuration (i.e., XY, $\gamma=\theta+90^{\circ}$), its intensity I(A$_{\text{1g}}$) = 0. These results are in agreement with the experimental data. Interestingly, the polarization profiles of CEFs Raman modes are very similar to the polarization profiles of E$_{\text{2g}}$  phonon modes, suggesting that CEFs modes share similar forms of Raman tensors to E$_{\text{2g}}$ phonon modes. 

\section{Polarization and Angular Dependence}

\begin{figure}[htbp]
\centering
\includegraphics[width=1.0\linewidth]{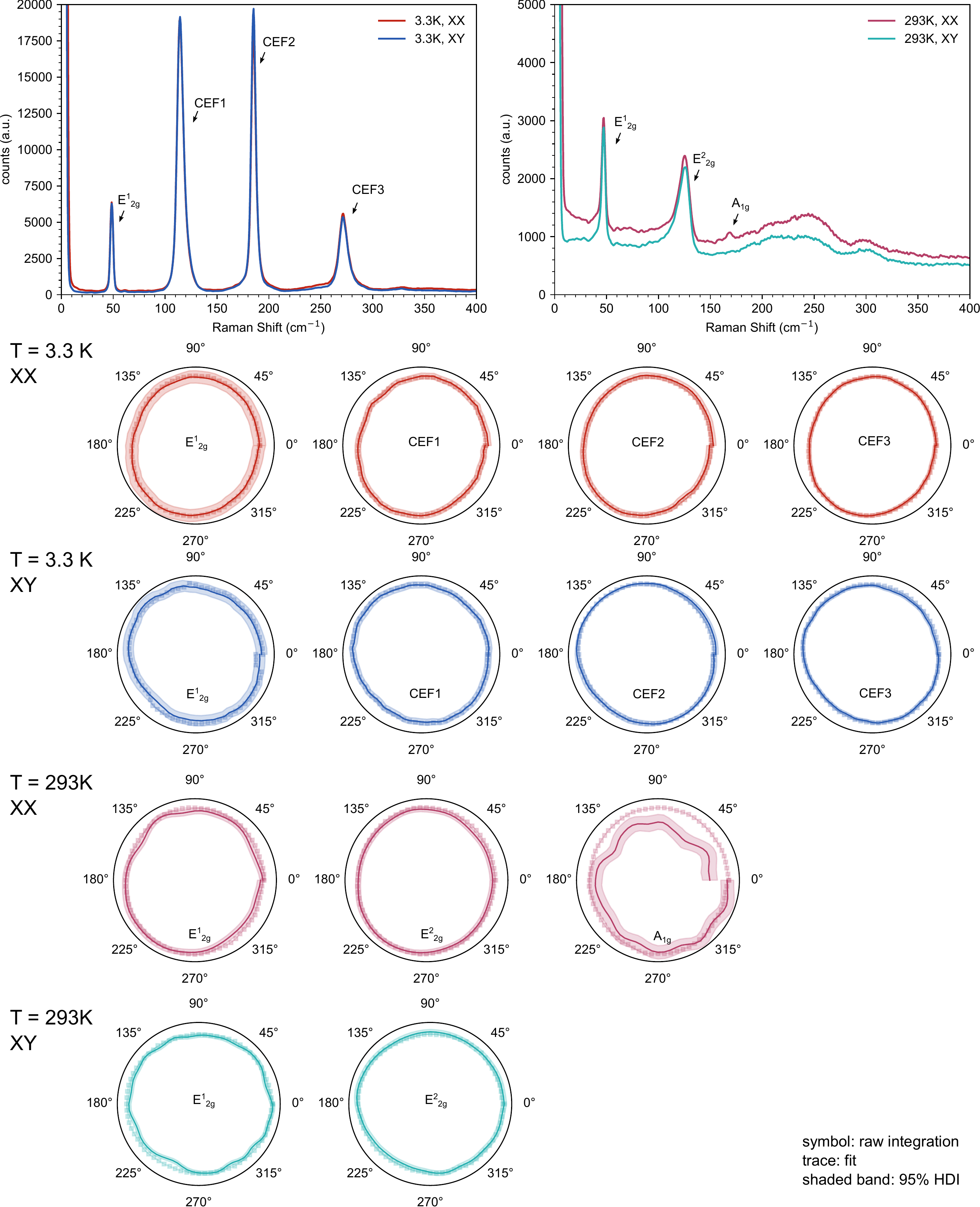}
\caption{Polarization and angular dependence of the at $T =$ 3.3 K and $T =$ 293 K.}
\label{fig:pol}
\end{figure}

Figure \ref{fig:pol} shows the polarization dependence of the peaks E$^1_{\text{2g}}$, CEF1, CEF2 and CEF3 $T =$ 3.3 K, E$^1_{\text{2g}}$, E$^2_{\text{2g}}$ and A$_{\text{1g}}$ at $T =$ 293 K. 

\newpage

\section{Calculated phonon dispersion}
\begin{figure}[htbp]
\centering
\includegraphics[width=0.5\linewidth]{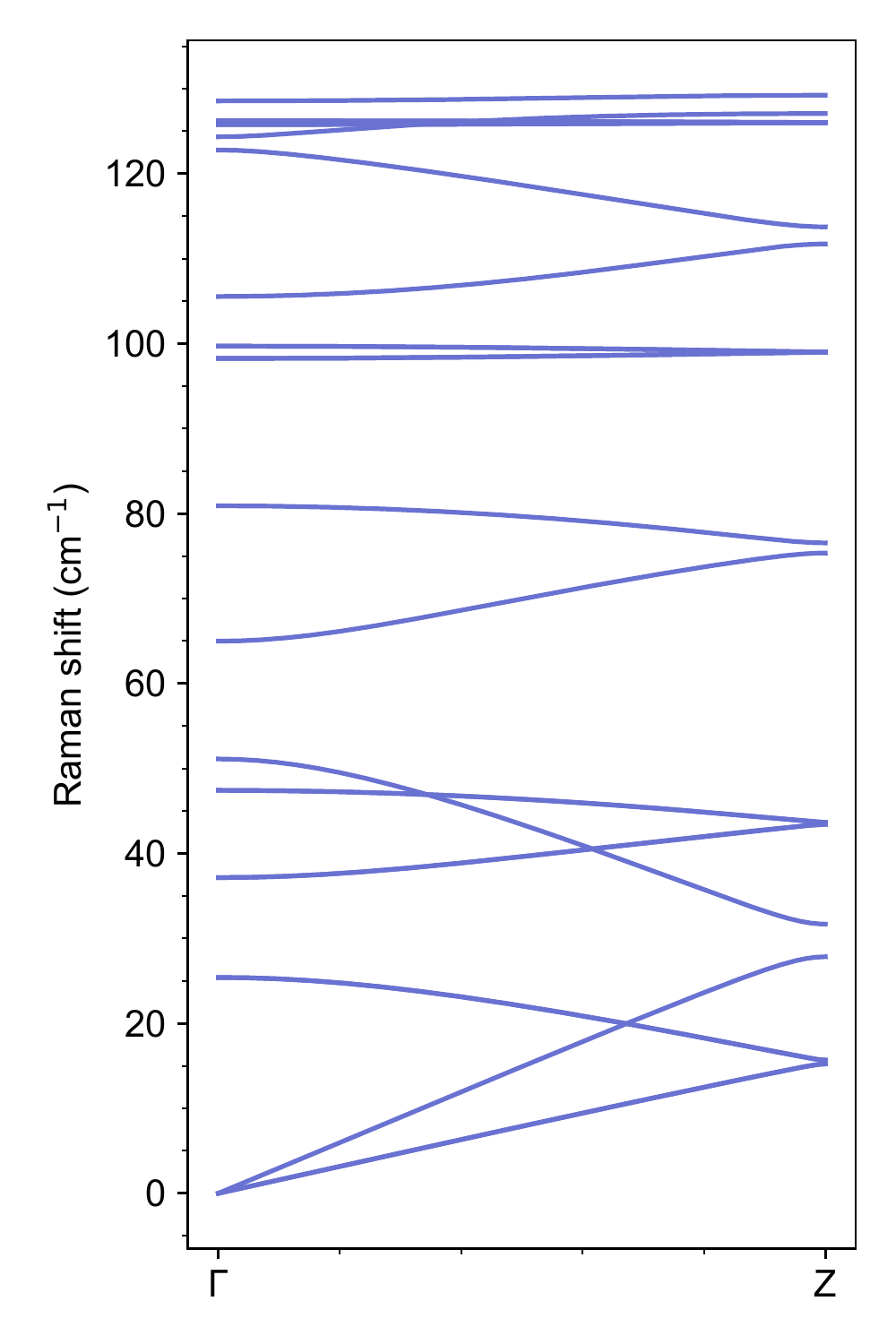}
\caption{Calculated phonon dispersion}
\label{fig:dis}
\end{figure}

\newpage
\section{Position Dependence}
\begin{figure}[htbp]
\centering
\includegraphics[width=1.0\linewidth]{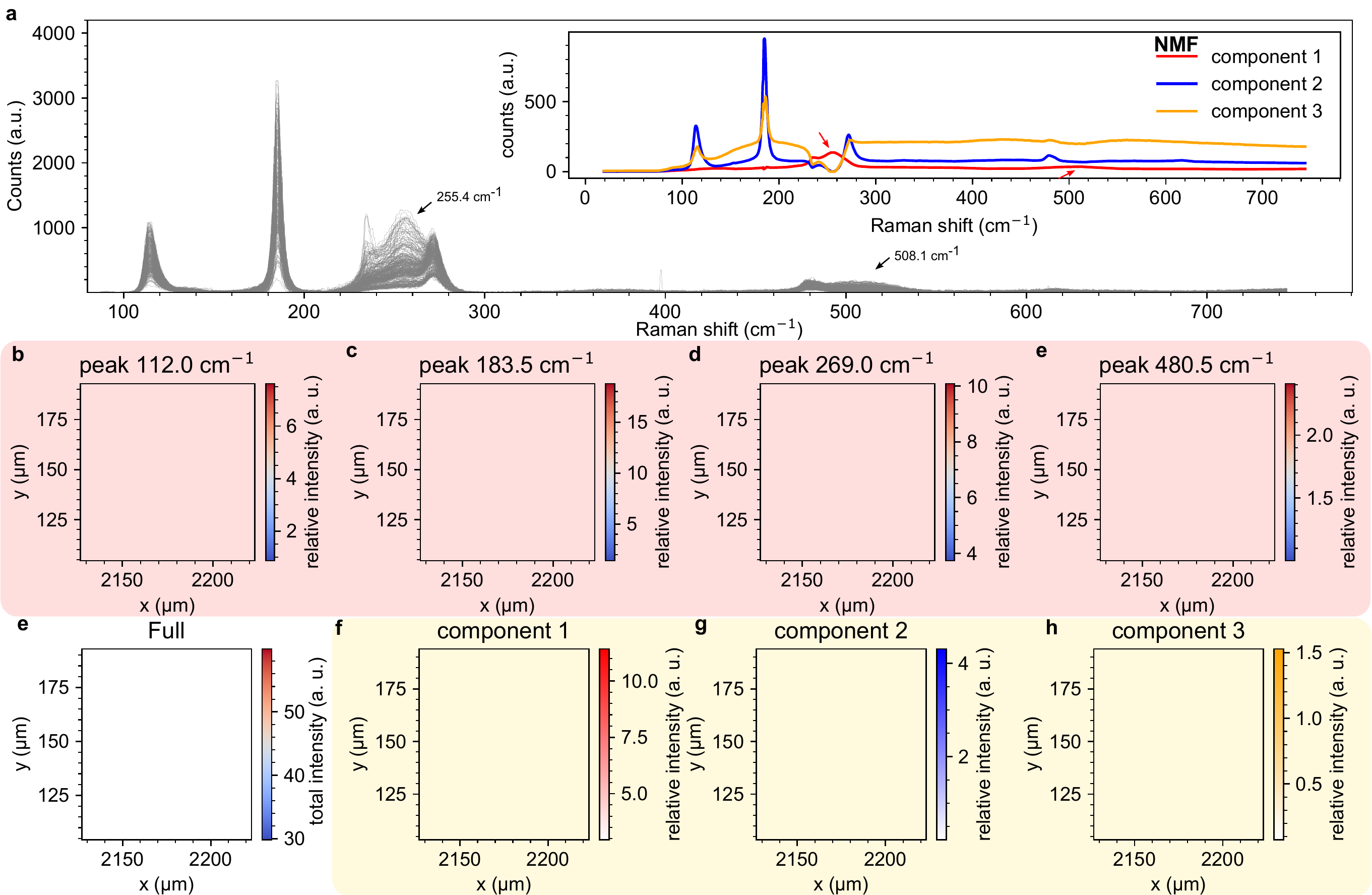}
\caption{Position Dependence at $T$ = 3.3 K.}
\label{fig:spatial_3K}
\end{figure}

\begin{figure}[htbp]
\centering
\includegraphics[width=1.0\linewidth]{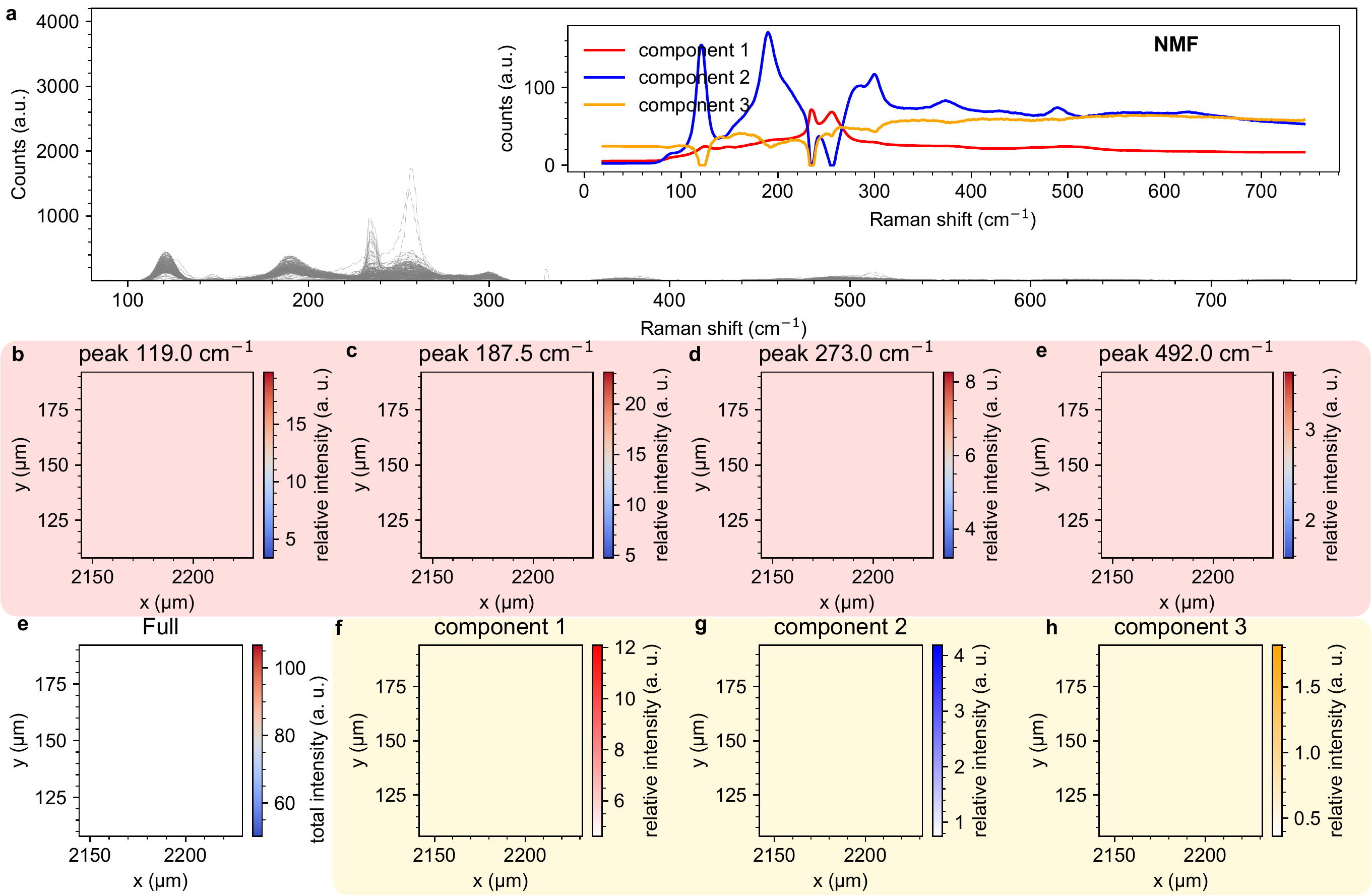}
\caption{Position Dependence at $T$ = 120 K.}
\label{fig:spatial_120K}
\end{figure}

\begin{figure}[htbp]
\centering
\includegraphics[width=1.0\linewidth]{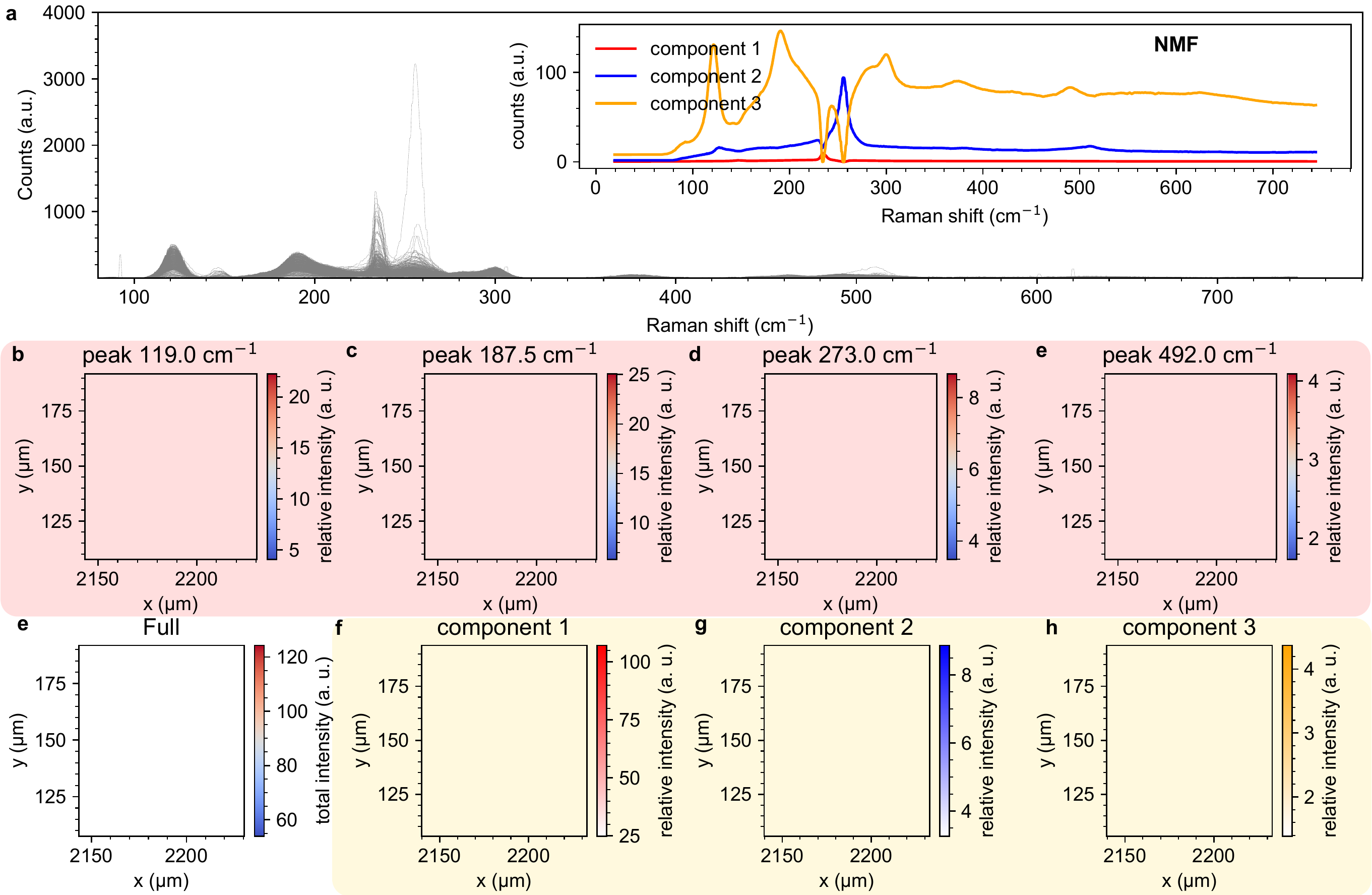}
\caption{Position Dependence at $T$ = 130 K.}
\label{fig:spatial_130K}
\end{figure}

As described in the main text, we report subtle spatial anti-correlations between phonon and CEF modes, such as the CEF1, E$^2_{\text{2g}}$, and $\omega_2$ modes. Simple spatial plots of integrated counts over a specific peak may be affected by baseline corrections and/or large peaks nearby. The baseline can be removed by asymmetric least squares fitting but the accuracy is not necessarily satisfactory. Meanwhile, full-blown curve fitting over thousands of spatial points can be computational expensive. Non-negative matrix factorization (NMF) is a simple algorithm that captures the most linearly independent basis vectors out of a given hyper-dimensional data cube with very low computational cost, yet it is effective in exploratory data analysis. Here Figures \ref{fig:spatial_3K}, \ref{fig:spatial_120K}, and \ref{fig:spatial_130K} illustrate spatially resolved Raman spectra at $T = 3$ K, $T = 120$ K, and $T = 130$ K, respectively. In these figures, (a) illustrates a subset of raw Raman spectra, with the inset illustrating the captured NMF components. (b-e) are integrated counts for selected peaks. (f-h) are the weights with respect to the basis vector components for the NMF decomposition. A similar anticorrelation between modes to that described in the main text is observed again in these representations. The $(x, y)$ are the raw coordinate values. A value of $(x, y) =(2127 \;\mu\text{m}, 104 \;\mu\text{m})$ was subtracted in the main text.

\subsection{T = 3K}

\newpage

\subsection{T = 120K}

\newpage

\subsection{T = 130K}

\newpage